\documentstyle[amssymb,12pt,epsfig]{article}
\begin{document}
\begin{center}
{\bf L.G. Mardoyan$^1$, L.S. Petrosyan$^2$, H.A. Sarkisyan$^3$}        
\end{center}

\begin{center}
{\large\bf The charge-dyon bound system in the spherical quantum well}
\end{center}

\begin{center}
Yerevan State University, \\
1, Alex Manoogian, 375025, Yerevan, Armenia \\ [1mm]
\end{center}

\vspace{0.5cm}

{\small The spherical wave functions of charge-dyon bounded system
in a rectangular spherical quantum dot of infinitely and finite
height are calculated. The transcendent equations, defining the
energy spectra of the systems are obtained. The dependence of the
energy levels from the wall sizes is found.}

\footnotetext[1]{e-mail: mardoyan@icas.ysu.am}
\footnotetext[3]{e-mail: lpetros@www.physdep.r.am}
\footnotetext[2]{e-mail: shayk@www.physdep.r.am}

It is known [1], that the charge-dyon system possesses the hidden symmetry,
and the group of hidden symmetry for discrete spectrum is $SO(4)$, and for
the continuous spectrum is $SO(3,1)$. Let's remind, that the dyon is a
hypothetical particle, introduced by Schwinger [2], which serves as a
source of both electrical and magnetic fields. The problem of
charge-dyon system due to the hidden symmetry can be factorized not
only in spherical but also in parabolic coordinates.

In the present article the charge-dyon bounded system is
considered in a spherical quantum dot of finite height. As a cause
of such investigation on the one hand served the recently
established fact (named dyon-oscillator duality). According to
this fact 4-dimensional isotropic oscillator is dual to the
3-dimensional charge-dyon system [3-5]. However, the property of
dyon-oscillator duality is inherent not only to ${\rm I \!R}^4 \to
{\rm I \!R}^3$ mapping, but to ${\rm I \!R}^1 \to {\rm I \!R}^1$,
${\rm I \!R}^2 \to {\rm I \!R}^2$ ¨ ${\rm I \!R}^8 \to {\rm I
\!R}^5$, mappings as well. In the first two cases one-dimensional
and two-dimensional anyons are obtained at the output [6, 7], in
third case non-Abelian Yang monopole arise [8].

On the other hand, due to the Hamiltonian of the system, factorable in the
same coordinates, as in the case of the hydrogen-like system, it becomes
possible to state that this Hamiltonian serves as a generalization of the
hydrogen-like system in case of magnetic charge. From this point of view the
considered problem is the generalization of the well- known problem of the
hydrogen-like system behavior inside semiconductor quantum dots with
different confinement potentials [9-13]. In other words it should be
expected, that in some cases energy levels of charge-dyon system are
analogical to hydrogen-like levels inside quantum dot. In this
connection let's mention that the behavior of the hydrogen-like system in a
rectangular spherical quantum dot of finite height has been
investigated in [9].

Let's consider the charge-dyon bounded system inside potential
sphere with confinement potential, which looks like
\begin{eqnarray*}
U(r) = \cases{0, \qquad  r < r_0; \cr
U_0, \qquad r \geq r_0. \cr}
\end{eqnarray*}
The Schr\"odinger equation of such system can be written as
\begin{eqnarray}
\left(\frac{\partial }{\partial x_j} - i
\frac{e}{\hbar c}A_j\right)^2 \psi
+ \frac{2\mu}{\hbar^2}\left[E + \frac{e^2}{r}
-\frac{\hbar^2s^2}{2\mu r^2} - U(r)\right]\psi = 0,
\end{eqnarray}
with the vector potential
\begin{eqnarray*}
{\vec A} = \frac{g}{r(r + z)}\left(y, - x, 0\right)
\end{eqnarray*}
which corresponds to the Dirac monopole [14] with the magnetic charge
$g=\frac{\hbar cs}{e}$ ($s=0,\pm \frac{1}{2},\pm 1,\dots$), and
singularity axis at $z > 0$.

We look for the solution of Eq.(1) in spherical coordinates as
\begin{eqnarray}
\psi(r,\theta,\varphi) = R(r)
Z(\theta)\,\frac{e^{im\varphi}}{\sqrt{2\pi}},
\end{eqnarray}
where $m=-|s|,|s|+1,\dots,|s|-1,|s|$. Now, inserting Eq.(2) into
Eq.(1), which is the Schr\"odinger equation in spherical
coordinates we come to the following pair of ordinary differential
equations
\begin{eqnarray}
\frac{1}{\sin \theta} \frac{d}{d \theta}\left(\sin \theta \frac{d
Z}{d \theta}\right) - \left[\frac{(m-s)^2}{2(1-\cos \theta)} +
\frac{(m+s)^2}{2(1+\cos \theta)}\right]Z + \ell(\ell+1)Z=0, \\
[3mm] \frac{1}{r^2}\frac{d}{dr}\left(r^2\frac{dR}{dr}\right) -
\frac{\ell(\ell+1)}{r^2}R + \frac{2\mu}{\hbar^2} \left(E +
\frac{e^2}{r} - U(r)\right)R = 0,
\end{eqnarray}
where $\ell(\ell+1)$ is a separation constant, and $\ell$ is a non-negative
integer or a half-integer which accepts the following values
\begin{eqnarray*}
\ell = \frac{|m-s|+|m+s|}{2},\frac{|m-s|+|m+s|}{2}+1,\dots.
\end{eqnarray*}
Fermion values of orbital moment are conditioned by the dyon magnetic charge.

The solution of Eq.(3) normalized by the condition
\begin{eqnarray*}
\int_{0}^\pi\,Z_{\ell m s}(\theta)Z_{\ell' m s}(\theta)
\sin\theta\,d\theta = \delta_{\ell \ell'}
\end{eqnarray*}
is the form
\begin{eqnarray*}
Z_{\ell m s}(\theta) = \sqrt{\frac{2\ell+1)}{4\pi}}
d^{\ell}_{ms}(\theta),
\end{eqnarray*}
where $d^{\ell}_{ms}(\theta)$ is the Wigner function [15].

For the radial equation (4) we obtain two solutions, which satisfy
standard conditions inside quantum dot ($r < r_0$) and outside it
($r \geq r_0$)
\begin{eqnarray}
R(r) = \cases{R^{(1)}_{k_1\ell}(r) =C^{(1)}_{k_1 \ell}
e^{-\gamma_1 r}r^{\ell}F\left(\ell +1 - k_1; 2\ell + 2; 2\gamma_1 r\right),
 \qquad  r < r_0; \cr
\cr
R^{(2)}_{k_2\ell}(r) = C^{(2)}_{k_2 \ell}e^{-\gamma_2 r}r^{\ell}
U\left(\ell +1 - k_2; 2\ell + 2; 2\gamma_2 r\right), \qquad r \geq r_0, \cr}
\end{eqnarray}
where
\begin{eqnarray*}
\gamma_1^2 = - \frac{2\mu E}{\hbar^2},\qquad
\gamma_2^2 = - \frac{2\mu (E -U_0)}{\hbar^2},\qquad
k_1^2 = -\frac{\mu e^4}{2\hbar^2 E}, \qquad
k_2^2 = -\frac{\mu e^4}{2\hbar^2 (E - U_0)},
\end{eqnarray*}
$F(\alpha; \beta; x)$ and $U(\alpha; \beta; x)$ are confluent
hypergeometrical functions of the first and the second kind, consequently.

Radial wave function normality condition can be written as
\begin{eqnarray}
\int_{0}^{r_0} r^2\,\left[R^{(1)}_{k_1 \ell}(r)\right]^2 dr +
\int_{r_0}^\infty r^2\,\left[R^{(2)}_{k_2 \ell}(r)\right]^2 dr
= 1.
\end{eqnarray}

From the continuity condition for the logarithmic derivative of
the radial wave functions $R^{(1)}_{k_1\ell}(r)$ and
$R^{(2)}_{k_2\ell}(r)$ in $r = r_0$ point
\begin{eqnarray}
\left(\frac{d\ln R^{(1)}_{k_1 \ell}}{dr}\right)_{r=r_0} =
\left(\frac{d\ln R^{(2)}_{k_2 \ell}}{dr}\right)_{r=r_0}
\end{eqnarray}
we obtain, that $C^{(2)}_{k_2 \ell}=C^{(1)}_{k_1 \ell}A_{k_1 k_2 \ell}$,
where
\begin{eqnarray*}
A_{k_1 k_2 \ell}= \frac{e^{(\gamma_2 -\gamma_1)r_0}
F\left(\ell +1 - k_1; 2\ell + 2; 2\gamma_1 r_0\right)}
{U\left(\ell +1 - k_2; 2\ell + 2; 2\gamma_2 r_0\right)}.
\end{eqnarray*}
Then, using the normality condition (6), for $C^{(1)}_{k_1 \ell}$ we obtain
the expression
\begin{eqnarray*}
C^{(1)}_{k_1 \ell} = \left[I^{(1)}_{k_1 \ell}+A^2_{k_1 k_2 \ell}
I^{(2)}_{k_2 \ell}\right]^{-1/2},
\end{eqnarray*}
where
\begin{eqnarray*}
I^{(1)}_{k_1 \ell} = \int_{0}^{r_0}
e^{-2\gamma_1 r}r^{2\ell + 2}
\left[F\left(\ell +1 - k_1; 2\ell + 2; 2\gamma_1 r\right)\right]^2dr,
\\ [2mm]
I^{(2)}_{k_2 \ell} = \int_{r_0}^\infty
e^{-2\gamma_2 r}r^{2\ell + 2}
\left[U\left(\ell +1 - k_2; 2\ell + 2; 2\gamma_2 r\right)\right]^2dr.
\end{eqnarray*}
Then wave functions (5) can be rewritten in the following form
\begin{eqnarray}
R(r) = \cases{R^{(1)}_{k_1\ell}(r) =C^{(1)}_{k_1 \ell}
e^{-\gamma_1 r}r^{\ell}F\left(\ell +1 - k_1; 2\ell + 2; 2\gamma_1 r\right),
 \qquad  r < r_0; \cr
\cr
R^{(2)}_{k_2\ell}(r) = C^{(1)}_{k_1 \ell}A_{k_1 k_2 \ell}
e^{-\gamma_2 r}r^{\ell}
U\left(\ell +1 - k_2; 2\ell + 2; 2\gamma_2 r\right), \qquad r \geq r_0. \cr}
\end{eqnarray}
Now, using the relationship (8) from the condition (7) we obtain a
transcendental equation, defining the spectrum of the charge-dyon bounded
system in the spherical quantum dot
\begin{eqnarray*}
\gamma_1 - \frac{2\gamma_1 (\ell +1 - k_1)
F\left(\ell +2 - k_1; 2\ell + 3; 2\gamma_1 r_0\right)}{(2\ell + 2)
F\left(\ell +1 - k_1; 2\ell + 2; 2\gamma_1 r_0\right)} = \nonumber \\
\\
= \gamma_2 + \frac{2\gamma_2 (\ell +1 - k_2)
U\left(\ell +2 - k_2; 2\ell + 3; 2\gamma_2 r_0\right)}
{U\left(\ell +1 - k_2; 2\ell + 2; 2\gamma_2 r_0\right)}. \nonumber
\end{eqnarray*}

Let us note that the spectrum of the charge-dyon system is
\begin{eqnarray*}
E_n = -\frac{\mu e^4}{2\hbar^2n^2},
\end{eqnarray*}
where $n=1,3/2,2\dots$ is a principal quantum number, and the
multiplicity of degeneration of the energy levels at fixed $n$
and $s$ is [5]
\begin{eqnarray*}
g^{s}_n = (n-s)(n+s).
\end{eqnarray*}

The curves of energy dependencies of the charge-dyon system in the
spherical infinitely high potential well via well radius $r_0$ (in
$E_R$ and $a_B$) for  $\ell = 0; 0.5; 1; 1.5$ cases are shown on
Fig. 1. As it follows from this figure, at small $r_0$ the
arrangement of the levels corresponds to the one, which takes
place when single particle falls into infinitely potential well.
The only difference is that besides the well-known levels, which
correspond to the integer values of $\ell$, intermediate levels
with fermion values of $\ell = 0.5; 1.5$ exist in this case. As is
visible from Fig. 1 at small $r_0$ for given $n_r=n-\ell -1$ they
are arranged between levels with integer $\ell$ as $\ell$
increases. At the increase in $r_0$ charge-dyon interaction
becomes more essential and levels pass to ones, which correspond
to the free charge-dyon system. In other words the level
degeneracy which corresponds to the charge-dyon free system is
step by step reestablished at the increase in $r_0$. In
particular, $2S$ level crosses the lowest level with $\ell = 1.5$
and joins $2P$ level on the same curve. In its turn the first
level with $\ell = 1.5$ joins the second level $\ell = 0.5$. At
that all these levels go down at increasing $r_0$.

On Fig.2 the same $E(r_0)$ dependence is shown for the case of the
well of finite height $(U_0=5E_R)$. In the case of large $r_0$ the
behavior of curves is analogous with the one, which takes place at
consideration of infinitely high well. As $r_0$ decreases at first
the influence of walls rise in analogy with the upper considered
case, that is why the levels are arranged as in infinitely high
spherical well. However, at too small values of $r_0$ the role of
the "first medium" $(r<r_0)$ becomes inessential. In accordance
with this fact the energy levels of the system pass to those,
which take place for the charge-dyon system, but shifted up by the
values $U_0$, i.e. $E \to U_0 - E_R/n^2$. It explains the joining
of curves (e.g. $2P$ and $2S$) at $r_0 \to 0$.

Thus,  the presence of competition between the energy of the
charge interaction with the well walls and energy of charge-dyon
interaction is typical for both cases. At comparatively small
radii of the well (we don't mean the values $r_0 \to 0$) the main
contribution in the energy of the system is conditioned by
repulsing potential of the well walls, that's why the levels are
positive. The charge-dyon interaction begins to play the main role
at increasing $r_0$ and the levels become negative.

On Fig.3 the curves of the observed system's energy dependence via
the height of the well $E(U_0)$ are presented (in $E_R$ units). As
it should be expected, at increasing $U_0$ the levels go up. At
that the lower levels, beginning from some $U_0$ almost don't feel
the enlargement of the well height.

Now let's consider the point, the analogical to which plays an
important role in the theory of impurity states in quantum dots.
Usually, in solid state problems along with the energy of impurity
-- quantum dot system one has to calculate the binding energy of
the impurity inside quantum dot as well. This energy is defined as
the difference between the energies of the free and impurity
electron inside the quantum dot [10]. Translating it to the
language of the problem under the consideration, it is necessary
to find the difference between the energies, in one case defined
by the Eq.(1), in other case defined by the analogical equation,
but without Coulomb interaction term $e^2/r$. For that we find
wave functions and energy spectrum of the equation
\begin{eqnarray}
\left(\frac{\partial }{\partial x_j} - i
\frac{e}{\hbar c}A_j\right)^2 \psi^{(0)}
+ \frac{2\mu}{\hbar^2}\left[E_0
-\frac{\hbar^2s^2}{2\mu r^2} - U(r)\right]\psi^{(0)} = 0.
\end{eqnarray}
The angular part of the Eq.(10) coincides with Eq.(3). For radial
part we come to the equation
\begin{eqnarray}
\frac{1}{r^2}\frac{d}{dr}\left(r^2\frac{dR^{(0)}}{dr}\right)
- \frac{\ell(\ell+1)}{r^2}R^{(0)} + \frac{2\mu}{\hbar^2}
\left(E_0 - U(r)\right)R^{(0)} = 0.
\end{eqnarray}
The solutions of the Eq.(10) inside and outside the well are given
by the following expressions
\begin{eqnarray*}
R^{(0)}(r) = \cases{R^{(1)}_{k_0\ell}(r) =C_{k_0\ell}
j_\ell(k_0r),  \qquad  r < r_0; \cr \cr R^{(2)}_{k\ell}(r) =
A_{k\ell}C_{k \ell}\frac{1}{\sqrt r} K_{\ell +\frac{1}{2}}(k r),
\qquad r \geq r_0, \cr}
\end{eqnarray*}
where $k_0= \sqrt{2\mu E/\hbar^2}$, $k=\sqrt{2\mu
(U_0-E)/\hbar^2}$, $j_\ell(x)$ is Bessel spherical function,
$K_{\ell +\frac{1}{2}}(x)$ is McDonald function, and
\begin{eqnarray*}
A_{k_0k\ell} = \frac{{\sqrt r_0} j_\ell(k_0 r_0)}
{K_{\ell+\frac{1}{2}}(kr_0)}, \qquad
C_{k_0k\ell}^2 = \frac{1}{I_{k_0\ell}+A_{k\ell}^2I_{k\ell}},
\nonumber \\ [3mm]
I_{k_0\ell} = \int_{0}^{r_0} r^2\,\left[j_\ell (k_0 r)\right]^2 dr, \qquad
I_{k\ell} = \int_{r_0}^{\infty} r^2\,\left[K_{\ell+\frac{1}{2}}(k r)
\right]^2 dr.
\end{eqnarray*}

From the continuity condition of logarithmical derivatives of
$R^{(1)}_{k_0\ell}$ and $R^{(2)}_{k\ell}$ in $r=r_0$ point we come
to the transcendent equation, which defines the energy spectrum
$E_0$ of an electron, described by the Eq.(9).
\begin{eqnarray*}
\frac{j'_\ell(k_0r_0)}{j_\ell(k_0r_0)} =
\frac{\left(\frac{1}{\sqrt r} K_{\ell +\frac{1}{2}}(kr_0)\right)'}
{\frac{1}{\sqrt r} K_{\ell +\frac{1}{2}}(kr_0)}.
\end{eqnarray*}

On Fig.4 the curves of electron energy $E_0$ dependencies via well
radius $r_0$ are shown for the values $\ell=0;0.5;1;1.5$. Let's
mention at once, that the appearence of levels has threshold
character, i.e. at increasing well height $U_0$ other new levels
arise. The arrangement of levels is analogical to the one, which
takes place for single particle in spherical well of finite depth.
As $r_0$ increases all levels come closer to each other and go
lower. Here due to the presence of magnetic charge besides the
well-known levels with integer values of $\ell$ intermediate
levels with half-integer values of $\ell$ arise as well.

Now, defining the analogy to the binding energy of charge-dyon
system in quantum dot as the $E_b=E_0-E$ difference, one can plot
$E_b(r_0)$ for different values of $\ell$. This dependence is
presented on Fig.5. According to this figure, at large $r_0$, when
the influence of well walls is inessential, $E_b$ levels by
absolute value tend to ones, which are typical for the free system
of charge-dyon. In particular the ground level (continuous curve)
tends to $E_R$ value. At decreasing $r_0$ the influence of walls
increases, which results in increasing levels. After the levels
achieved their maximal value, the levels begin to go down, since
the influence of well walls at to small radii begins to weaken. It
looks like the particle jumps out from the well. Let's mention,
that the analogical situation is observed in solid state problems
(see, e.g. [16, 17]).

\vspace{1cm}

\noindent {\bf Acknowledgments}

We would like to thank Dr. Armen Nersessian for useful
discussions.

The research of L.S. Petrosyan and H.A. Sarkisyan supported by
INTAS grants No. 99-00928 and 0175WP.

The work of L.G. Mardoyan supported by the ANSEF grant No. PS81.

\begin{figure}[tbph]
\label{fig1}\epsfig{figure=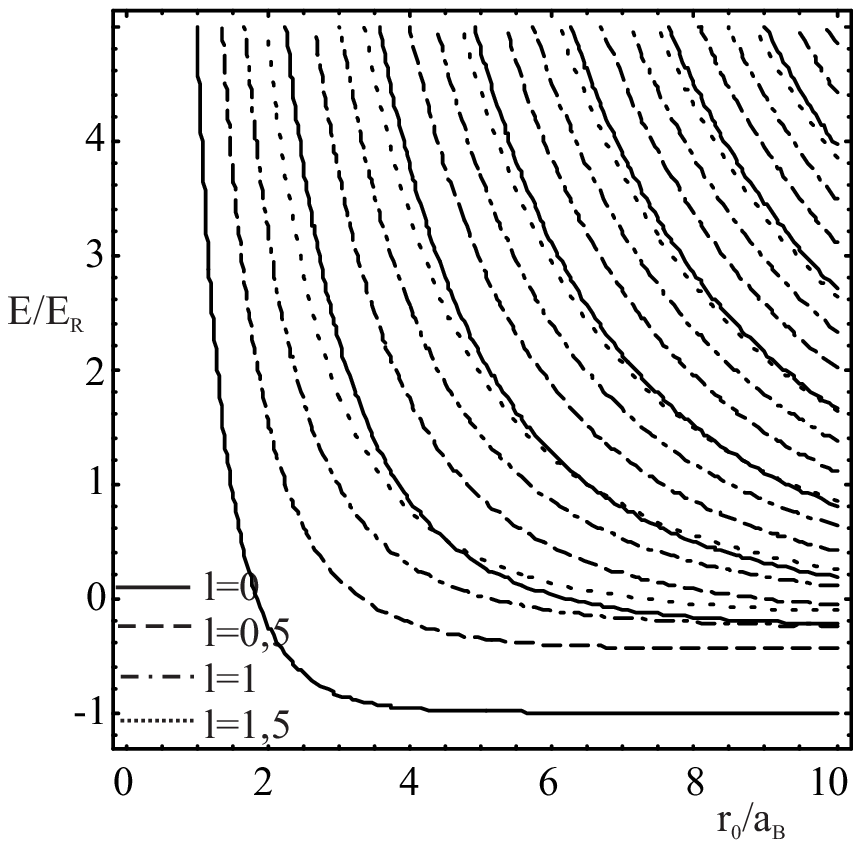,width=7cm,height=7cm}~~~~~~~~~ %
 \epsfig{figure=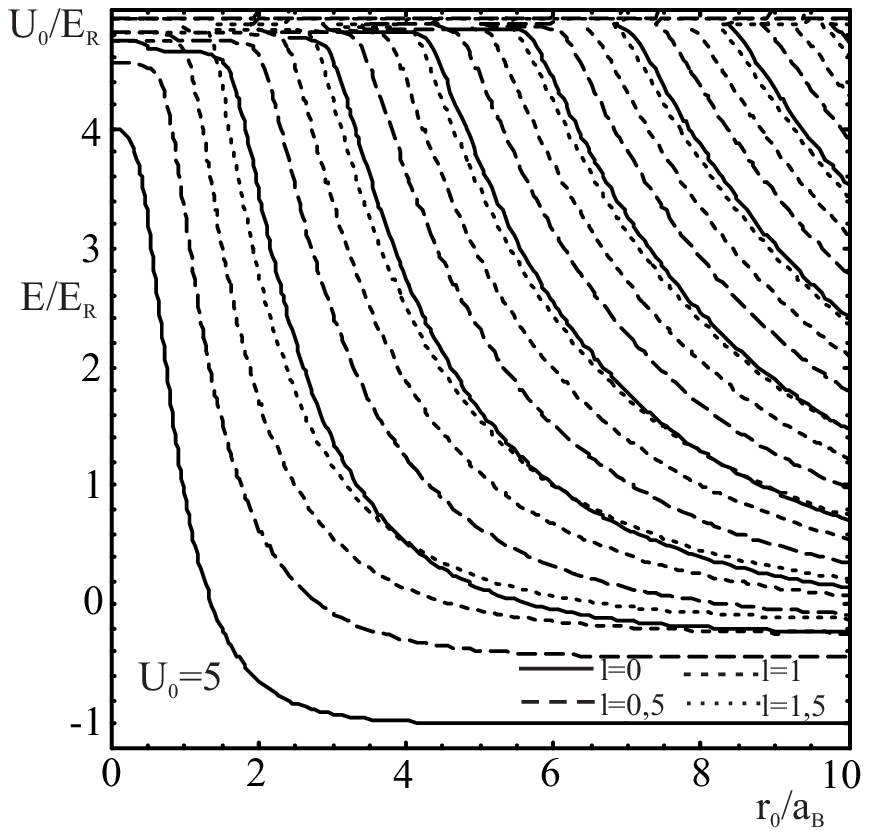,width=7cm,height=7cm}
\caption{The dependence of the full energy of the charge-dyon
system via the radius of infinitely high quantum dot for $\ell =
0;0.5;1;1.5$ cases (left figure).}\caption{The dependence of the
full energy of the charge-dyon system via the radius of quantum
dot of finite height for $\ell = 0;0.5;1;1.5$, $U_{0}=5E_{R}$
cases (right figure).} \label{fig2}
\end{figure}

\begin{figure}[tbph]
\epsfig{figure=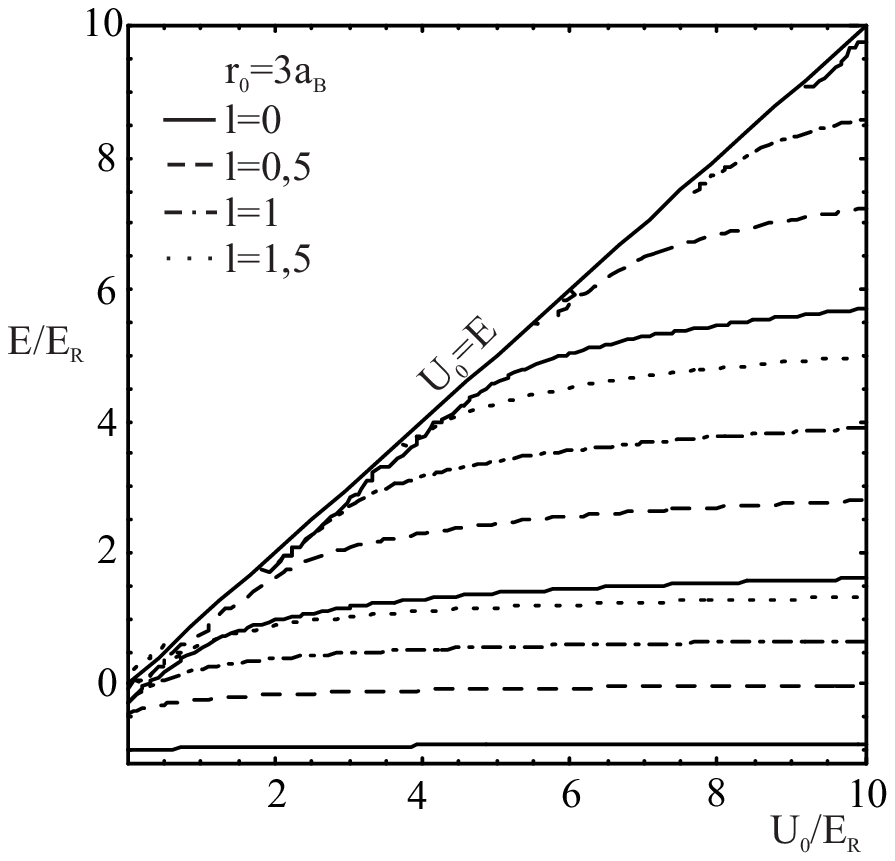,width=7cm,height=7cm}
\label{fig3}~~~~~~~~~%
\epsfig{figure=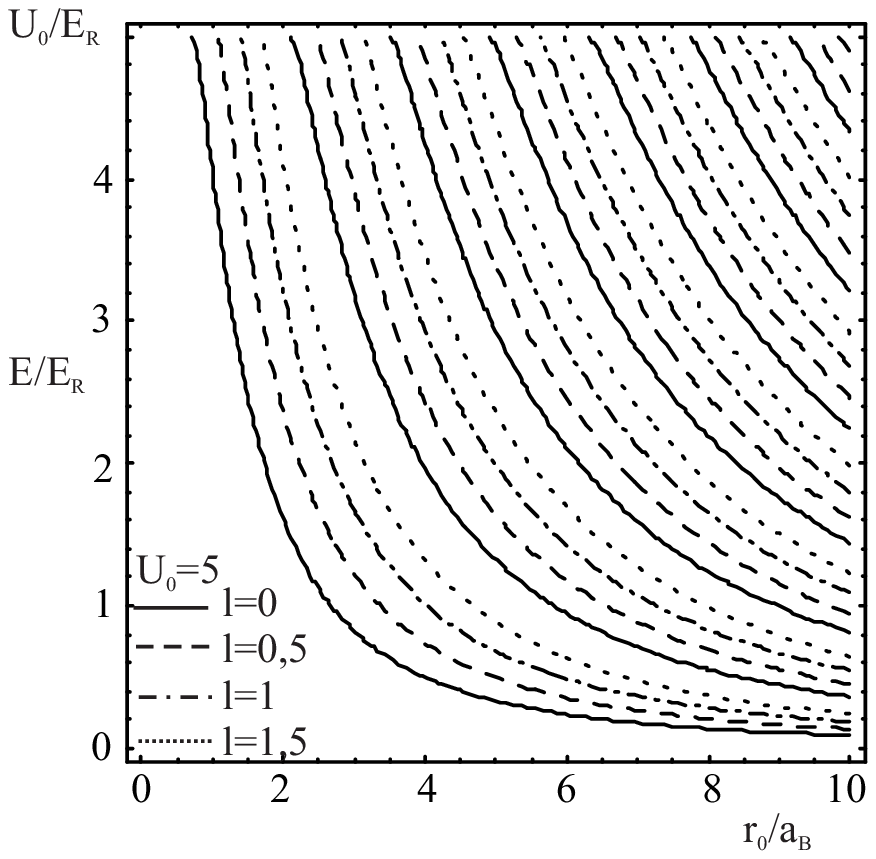,width=7cm,height=7cm}\caption{The
dependence of the full energy of the charge-dyon system via the
height of quantum dot $U_{0}$ for $\ell = 0;0.5;1;1.5$,
$r_{0}=3a_{B}$ cases (left figure).} \caption{The dyon energy
dependence via the radius of quantum dot for $\ell = 0;0.5;1;1.5$,
$U_{0}=5E_{R}$ cases (right figure).} \label{fig4}
\end{figure}

\begin{figure}[tbph]
\begin{center}
\epsfig{figure=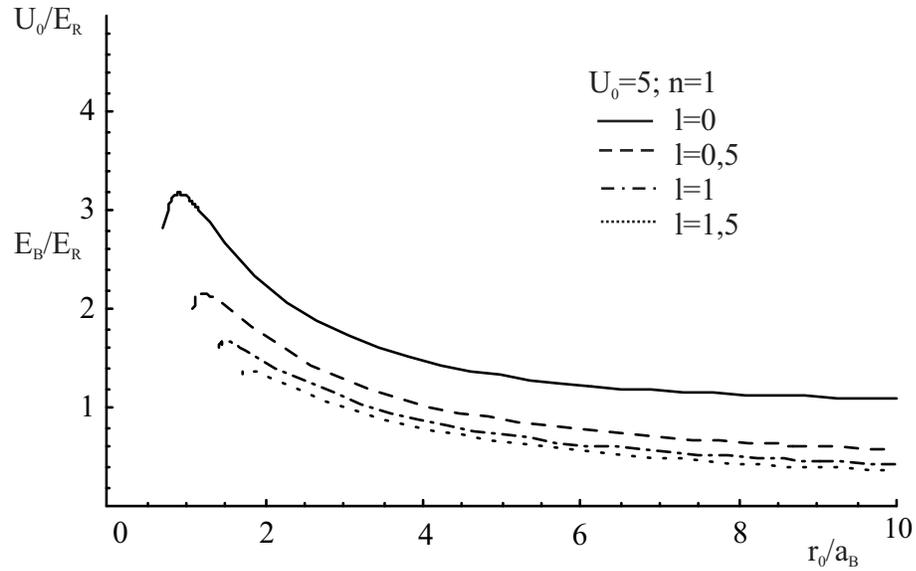,width=12cm,height=7.5cm} \caption{The
charge-dyon system's binding energy dependence via the radius of
quantum dot for $\ell = 0;0.5;1;1.5$, $U_{0}=5E_{R}$, $n=1$
cases.} \label{fig5}
\end{center}
\end{figure}


\vspace{2cm}

\begin{thebibliography}{99}
\bibitem{1}
D. Zwanziger, {\it Phys. Rev.} {\bf 176}, 1480, (1968).
\bibitem{2}
J. Schwinger, {\it Science} {\bf 165}, 757, (1969).
\bibitem{3}
A. Nersessian, V. Ter-Antonyan, {\it Mod. Phys. Lett}.
{\bf A9}, 2431, (1994).
\bibitem{4}
A. Nersessian, V. Ter-Antonyan, {\it Mod. Phys. Lett}. {\bf A10},
2633, (1995).
\bibitem{5}
L.G. Mardoyan, A.N. Sissakian, V.M. Ter--Antonyan,
{\it Int. J. Mod. Phys}. {\bf A12}, 237, (1997).
\bibitem{6}
Ye. Hakobyan, V. Ter-Antonyan. {\it Quantum oscillator as 1D anyon}.
quant-ph/0002069.
\bibitem{7}
A. Nersessian, V. Ter-Antonyan, M. Tsulaia {\it Mod. Phys. Lett.}
{\bf A19}, 1605, (1996).
\bibitem{8}
L.G. Mardoyan, A.N. Sissakian, V.M. Ter-Antonyan,
{\it Oscillator as a Hidden Non--Abelian Monopole}. Hep--th/9601093.
\bibitem{9}
J.-L. Zhu, {\it Phys. Rev.} {\bf B39}, 8780, (1989).
\bibitem{10}
D.S. Chuu, C.M. Hsiao, W.N. Mei, {\it Phys. Rev.} {\bf B46}, 3898, (1992).
\bibitem{11}
C. Bose, C. Sarkar, {\it Physica} {\bf B253}, 238, (1998).
\bibitem{12}
E.M. Kazaryan, L.S. Petrosyan, H.A. Sarkisyan, {\it Int. J. Mod. Phys.}
{\bf B15}, 4103, (2001).
\bibitem{13}
E.M. Kazaryan, L.S. Petrosyan, H.A. Sarkisyan, {\it Physica} {\bf E 16},
174, (2003).
\bibitem{14}
P.A.M. Dirac, {\it Proc. Roy. Soc}. {\bf A133}, 69, (1931).
\bibitem{15}
D.A. Varshalovich, A.N. Moskalev, and V.K. Khersonskii. Quantum
Theory of Angular Momentum. World Scientific, Singapore, (1988).
\bibitem{16}
C. Bose, C. Sarkar, {\it Phys. Stat. Sol.} {\bf B 218},
461, (2000).
\bibitem{17}
N. Porras-Montenegro, S. Perez-Merchancano, A. Latge, {\it J.
Appl. Phys.} {\bf 74}, 7624, (1993).
\end{thebibliography}
\end{document}